\documentclass[a4paper,12pt]{article}
\usepackage[english]{babel}
\usepackage {amssymb,amsmath,latexsym,enumerate,graphicx,epsfig,cite}
\usepackage{indentfirst}%абзацный отступ в первой строке
\usepackage{titlesec}%правка оформления заголовков секций
\usepackage {subfigure,epsfig}

\usepackage{caption2}% заменяем для pиcyнков ':' поcле номеpа pиcyнка на '.'
\makeatletter\renewcommand{\@biblabel}[1]{#1.} \makeatother
\usepackage{amsthm} %пакет для работы с окружениями типа Теорема
\begin{document}
\title{The modified simplest equation method to look for exact solutions of nonlinear partial differential equations.}
\author{%%Кудряшов Н.А.,
Efimova O.Yu.}
\date{} \maketitle
\section*{Introduction.}
Nonlinear differential equations and its systems are used to describe various processes in physics, biology, economics and so on. In nineteenth century it was shown that nonlinear equations are significant as much as linear ones, therefore there was the great progress of relevant mathematical theories, including the theory of nonlinear differential equations.

Analytical solutions allow to obtain important patterns of model relationships, features of solutions behavior at different initial and border conditions.  Exact solutions allow to test numerical calculations by one of the most reliable ways.

There are a lot of methods to look for exact solutions of nonlinear differential equations:
the inverse scattering transform \cite{Gardner1967, Ablowitz1973}, Hirota method \cite{Hirota1971, Kudryashov2004en}, the Backlund transform, the truncated Painleve expansion \cite{Weiss1983, Kudryashov1992, Kudryashov1993, Efimova2004en}.
Here we present the generalization of modified simplest equation method, that was proposed to look for exact solutions of ordinary differential equations \cite{Kudryashov2005, Kudryashov2008b}.  Note, that one of advantages of this method is possibility of its complete automatization.

The structure of the paper is as follows. In section 1 the modification of simplest equation method to look for exact solutions of nonlinear partial differential equations is presented. Using this method we obtain exact solutions of generalized Korteweg-de Vries equation with cubic source (section 2) and exact solutions of third-order Kudryashov-Sinelshchikov equation describing nonlinear waves in liquids with gas bubbles (section 3).

\section{The simplest equation method to look for exact solutions of PDEs.}

The simplest equation method to look for exact solutions of ordinary differential equations is based on two main ideas: 1) on the usage of simplest equation with known solutions and 2) on the analysis of possible singularity orders of solutions \cite{Kudryashov2004en, Kudryashov2005, Kudryashov2007, Kudryashov2008a, Kudryashov2008b, Efimova2009}. In papers \cite{Vitanov2010a, Vitanov2010b, Vitanov2010c} the simplest equation method is used to look for traveling-wave solutions of partial differential equations.

Here we present the method analogous to modified simplest equation method, which allows to obtain the exact solutions of nonlinear PDEs that do not belong to the traveling-wave solutions class.

We are looking for exact solutions of nonlinear PDEs of polynomial form
\begin{equation}
P[u,u_t,\ldots, u_t^{(k)}, u_x, \ldots u_{x}^{(l)}]=0
\end{equation}

We suggest that function $u(x,t)$ can be presented via the solution of <<simplest>> equation $Z$. As <<simplest>> equation we use linear differential equations
\begin{align}
&Z_{x}^{(l)}=b_1 Z_x^{(l-1)}+\ldots+b_{l-2} Z_x + b_{l-1} Z  \label{e:trvchp_prost_ur1} \\
&Z_t=c_1 Z_x^{(l-1)} + \ldots+ c_{l-2} Z_x+c_{l-1} Z \label{e:trvchp_prost_ur2}
\end{align}

Sequentially the modified simplest equation method to look for exact solutions of nonlinear partial equation can be described in the following way.

\textit{Step 1.} Determination of  possible orders of poles and zeros of traveling-wave solutions of equation studied.
  The singularity orders can be determined by step-by-step examination of leading terms or by power geometry methods \cite{Bruno2004a, Kudryashov2004en}.

\textit{Step 2.} Solution of PDE represents as function of  <<simplest equation>> solution taking into account orders of poles and zeroes, that have been determined at step 1
 \begin{equation}\label{e:trvchp_podst_ob_v}
u(x,t)=  \sum_i {a_{i} f_{i}(Z)}
\end{equation}
For example, if traveling-wave solution allows first-order pole, we can use the form
 \begin{equation}
u(x,t)=  a_0+a_1 \frac{Z_x}{Z}
\end{equation}
For $n$-order zero it's rational to use transformation
\begin{equation}
u(x,t)=  a_0+a_1 Z +\ldots + a_n Z^n
\end{equation}
and so on.

\textit{Step 3.} Substituting \eqref{e:trvchp_podst_ob_v} into the initial equation and taking into account relations \eqref{e:trvchp_prost_ur1} -- \eqref{e:trvchp_prost_ur2} and setting coefficient at different powers of $Z(x,t), \ldots, Z(x,t)_x^{l-1}$ equal to zero, we obtain the algebraic system. Its solution determine coefficients $a_{i}$ and parameters of simplest equations $b_j$, $c_j$ ($l=1\ldots l-1$). If it is necessary, the restrictions  at parameters of initial equation can be done also.

\textit{Step 4.} Solution of equation \eqref{e:trvchp_prost_ur1} determines subject to restrictions, obtained at step 3, within $l$ arbitrary functions of $t$, which by-turn are to be defined from equation \eqref{e:trvchp_prost_ur2}. Solution of original PDE expresses via solution of system \eqref{e:trvchp_prost_ur1}--\eqref{e:trvchp_prost_ur2} according to transformation \eqref{e:trvchp_podst_ob_v}.

Note, this method can be completely automated by means of symbolic computations environment \textit{Maple}.

\section{Exact solutions of generalized Korteweg-de Vries equation with cubic source.}
As a first example let us consider the generalized Korteweg–de Vries equation with cubic source
\begin{equation}
u_t+3u^2u_x + 3(uu_x)_x+u^3+u_{xxx}+p_1u(u^2+p_2u+p_3)+p_4u_x+p_5u_{xx}+p_6uu_x=0
\end{equation}
The travelling-wave solutions of this equation allow first-order pole, so we are looking for exact solutions in the form
\begin{equation} \label{e:trvchp_p1_podst1}
u(x,t)=a_1 \frac{z_x}{z}, \qquad z\equiv z(x,t)
\end{equation}
where
\begin{align} \label{e:trvchp_p1_podst2}
z_{xxx}=b_1 z_{xx} + b_2 u_x +b_3 u\\
z_t=c_1 z_{xx}+c_2 z_x \label{e:trvchp_p1_podst3}
\end{align}

Substituting transformation \eqref{e:trvchp_p1_podst1} into original equation and using relations  (\eqref{e:trvchp_p1_podst2}--\eqref{e:trvchp_p1_podst3}), we obtain the algebraic system.
Its solution gives us
1) restriction on coefficient of original equation
\begin{equation}
p_6 = p_1+2 p_5
\end{equation}
and 2) parameters of relations (\ref{e:trvchp_p1_podst1}-\ref{e:trvchp_p1_podst3}):
\begin{equation}
a_1 = 1, c_2 = p_1 p_2+p_3-p_4, c_1 = p_1+p_2-p_5, b_1 = -p_2, b_2 = -p_3, b_3 = 0
\end{equation}

The type of solutions depends on sign $p_2^2-4p_3$, namely
\begin{itemize}
\item
at $p_2^2-4p_3>0$

\begin{align*}
&u(x,t) = \frac{C_1 \lambda_1 \exp{(\lambda_1 \tilde{x})} + C_2  \lambda_2 \exp{(\lambda_2 \tilde{x})}}{C_0\exp{((p_1 +p_2- p_5)p_3 t)}+C_1 \exp{(\lambda_1 \tilde{x})} + C_2 \exp{(-\lambda_2 \tilde{x})}}
\\
&\tilde{x} =x + (p_2 p_5-p_2^2+p_3-p_4)t, \qquad \lambda_{1,2}= \frac{-p_2 \pm \sqrt{p_2^2-4 p_3}}{2}
\end{align*}
\item
at $p_3=p_2^2/4$
\begin{align*}
&u(x,t) = \frac{\lambda C_1+C_2(1+\lambda \bar x)}{C_0\exp(\lambda \tilde{x})+C_1+C_2\bar{x}}
\\
&\bar{x} =x+( p_2 p_5-p_4-3\lambda^2)t,\quad
\tilde{x} =x - (\lambda \left(p_1+ p_5+\lambda\right)+p_4)t,
\quad \lambda=-\frac{p_2}{2}
\end{align*}

\item
at $p_2^2-4p_3<0$

\begin{align*}
u(x,t)& = \frac{(\lambda_2 C_1 - \lambda_1 C_2) \sin{(\lambda_1 \tilde{x})} + (\lambda_1 C_1 +\lambda_2 C_2) \cos{(\lambda_1 \tilde{x})}}{C_0\exp{(-\lambda_2 x +(p_1 +p_2- p_5)p_3 t)}+C_1 \sin(\lambda_1\tilde{x})+ C_2 \cos{(\lambda_1 \tilde{x})}}
\\
\tilde{x}& =x + (p_2 p_5-p_2^2+p_3-p_4)t,
\qquad \lambda_1 = \frac{\sqrt{4 p_3-p_2^2 }}{2}\qquad \lambda_2=\frac{-p_2}{2}
\end{align*}
\end{itemize}
Here $C_1$, $C_2$ and $C_3$ are the arbitrary constants, non-zero one of which can be canceled.

\section{Exact solutions of third-order equation describing nonlinear waves in liquids with gas bubbles.}
As the next example we consider the Kudryashov-Sinelshchikov equation for describing the pressure waves in liquid
with gas bubbles \cite{Kudryashov2010}

\begin{equation} \label{e:trs_eq_bubbles}
u_t+\alpha u u_x +u_{xxx}-(u u_{xx}+u u_x)_x-\beta u_x u_{xx}-\sigma u_{xx}=0
\end{equation}

Substituting transformations
\begin{align}
\\u(x,t)=a_0+ z^n, \qquad z\equiv z(x,t)
\\z_{xx}= b_1 z_x+ b_2 z, \qquad
z_t=c_1z_x+c_2z
\end{align}
into the original equation \eqref{e:trs_eq_bubbles} and equating coefficients at different powers of $z$ and $z_x$, we obtain that the modified simplest equation method results in exact solutions of equation \eqref{e:trs_eq_bubbles}, if one of following conditions holds:
\begin{align}
\beta = -1, n = 2, b_1 = -1, b_2 = \frac{\alpha}{4}, a_0 = \frac{\sigma+3}{2}, c_1 = -\alpha-1-\sigma,
c_2 = \frac{\alpha (1+\sigma)}{4} \label{e:trchp_b_usl1}
\\
\sigma = -1, n = \frac{2}{\beta+2}, b_1 = -1, b_2 = \frac{\alpha(2+\beta)}{4}, a_0 = 1, c_1 = -\alpha, c_2 = 0 \label{e:trchp_b_usl2}
\\
\alpha = -\frac{\beta}{(1+\beta)^2}, n = 1, b_1 = -\frac{1}{1+\beta}, b_2 = 0, c_1 = -\frac{\sigma+\sigma \beta+1}{(1+\beta)^2}, c_2 = 0 \label{e:trchp_b_usl3}
\\
\sigma = -1, \beta = 2, n = 1/2, a_0 = 1, b_1^2+b_1-\alpha = 0, b_2 = 0, c_1 = -\alpha, c_2 = 0 \label{e:trchp_b_usl4}
\\
\beta = 0, n = 1, b_1 = -1, b_2 = \frac{\alpha}{2}, c_1 = -\frac{\alpha }{2}(1+a_0)-\sigma-1, c_2 = \frac{\alpha}{2}  (1+\sigma) \label{e:trchp_b_usl5}
\end{align}

Let us discuss resulting exact solutions at condition \eqref{e:trchp_b_usl1} in detail.
Functions  $u(x,t)$ are $z(x,t)$ are to satisfy conditions
\begin{equation}
\begin{aligned}
u(x,t) = (\sigma+3)/2+z(x,t)^2\\z_t = (-\alpha-1-\sigma)z_x+ \alpha (\sigma+1) z/4
\\
z_{xx} = -z_x+\alpha z/4
\end{aligned}
\end{equation}
So we obtain solutions of original equation \eqref{e:trs_eq_bubbles}
\begin{itemize}
\item
at  $\beta=-2$ и $\alpha>-1$
\begin{equation}
u(x,t)=\frac{\sigma+3}{2}+\exp{(z_1)}\left( C_1\exp{z_2}+C_2\exp{(-z_2)}\right)^2
\end{equation}
\item
at $\beta=-2$ и $\alpha=-1$
\begin{equation}
u(x,t)=\frac{\sigma+3}{2}+\exp{(z_1)}(C_1+C_2(x-\sigma t))^2
\end{equation}
\item
at $\beta=-2$ и $\alpha<-1$
\begin{equation}
u(x,t)=\frac{\sigma+3}{2}+\exp{(z_1)} \left( C_1\sin{z_2}+C_2\cos{z_2}\right)^2
\end{equation}
\end{itemize}
where we use notation
\begin{equation}
z_1= \left(1+\frac{3}{2} \alpha+\sigma+ \frac{1}{2}\alpha\sigma\right) t-x ,\qquad
z_2= \frac{1}{2}\sqrt{|\alpha+1|}(\alpha t+t+\sigma t-x)
\end{equation}
Here $C_1$  and $C_2$ are the arbitrary constants.

Solutions at conditions  \eqref{e:trchp_b_usl2}, \eqref{e:trchp_b_usl3}, \eqref{e:trchp_b_usl4} or \eqref{e:trchp_b_usl5} can be presented similarly.
Note, that $c_2=0$ corresponds to the travelling-wave solutions.

%\begin{figure}[t]
%\center%
%\epsfig{file=trs-2-3.eps,width=70mm,angle=-90}
%\caption{}
%\label{fig:trs-2-3}
%\end{figure}

\section*{Conclusion.}

The modification of simplest equation method to look for exact solutions of PDEs is presented.
The efficiency of this method was demonstrated by two different examples.  Obtained solutions of the third-order Kudryashov-Sinelshchikov equation for describing the pressure waves in liquid
with gas bubbles seem to be new.

\bibliographystyle{unsrt}    
\bibliography{my_refs}

\begin{thebibliography}{10}

\bibitem{Gardner1967}
C.S. Gardner, J.M. Greene, M.D. Kruskal, and R.M. Miura.
\newblock Method for solving the {K}orteweg-de {V}ries equation.
\newblock {\em Phys. Rev. Lett.}, 19:1095--1097, 1967.

\bibitem{Ablowitz1973}
M.J. Ablowitz, D.J. Kaup, A.C. Newell, and H.~Segur.
\newblock Method for solving the {S}ine-{G}ordon equation.
\newblock {\em Phys. Rev. Lett.}, 30:1262--1264, 1973.

\bibitem{Hirota1971}
R.~Hirota.
\newblock Exact solution of the {K}orteweg-de {V}ries equation for multiple
  collisions of solitons.
\newblock {\em Phys. Rev. Lett.}, 27:1192--1194, 1971.

\bibitem{Kudryashov2004en}
N.~A. Kudryashov.
\newblock {\em Analytical theory of nonlinear differential equations}.
\newblock Moscow-Izhevsk: Institute of Computer Investigations, 2004.
\newblock In Russian.

\bibitem{Weiss1983}
J.~Weiss, M.~Tabor, and G.~Carnevale.
\newblock The {P}ainlev\'e property for partial differential equations.
\newblock {\em J. Math. Phys.}, 24(3):522--526, 1983.

\bibitem{Kudryashov1992}
N.~A. Kudryashov.
\newblock Partial differential equations with solutions having movable
  first-order singularities.
\newblock {\em Phys. Lett. A}, 169(4):237--242, 1992.

\bibitem{Kudryashov1993}
N.~A. Kudryashov.
\newblock Truncated expansions and nonlinear integrable partial differential
  equations.
\newblock {\em Physics LettersA}, 178:99--104, 1993.

\bibitem{Efimova2004en}
O.~Yu. Yefimova and N.~A. Kudryashov.
\newblock Exact solutions of the {B}urgers-{H}uxley equation.
\newblock {\em Journal of Applied Mathematics and Mechanics}, 68(3):413--420,
  2004.

\bibitem{Kudryashov2005}
N.~A. Kudryashov.
\newblock Simplest equation method to look for exact solutions of nonlinear
  differential equations.
\newblock {\em Chaos, Solitons and Fractals}, 24(5):1217--1231, 2005.

\bibitem{Kudryashov2008b}
N.~A. Kudryashov and N.~B. Loguinova.
\newblock Extended simplest equation method for nonlinear differential
  equations.
\newblock {\em Applied Mathematics and Computation}, 2008.
\newblock In press.

\bibitem{Kudryashov2007}
N.~A. Kudryashov and M.~V. Demina.
\newblock Polygons of differential equations for finding exact solutions.
\newblock {\em Chaos, Solitons \& Fractals}, 33(5):1480--1496, 2007.

\bibitem{Kudryashov2008a}
N.~A. Kudryashov and N.~B. Loguinova.
\newblock Be careful with the {E}xp-function method.
\newblock {\em Communications in Nonlinear Science and Numerical Simulation},
  2008.
\newblock In press.

\bibitem{Efimova2009}
Olga~Yu. Efimova.
\newblock Exact solutions of systems describing interacting populations with
  low critical densities.
\newblock {\em Applied Mathematics and Computation}, 208:134--143, 2009.

\bibitem{Vitanov2010a}
N.~K. Vitanov.
\newblock Application of simplest equations of {B}ernoulli and {R}iccati kind
  for obtaining exact traveling-wave solutions for a class of {PDE}s with
  polynomial nonlinearity.
\newblock {\em Communications in Nonlinear Science and Numerical Simulation},
  15:2050--2060, 2010.

\bibitem{Vitanov2010b}
N.~K. Vitanov, Z.~I. Dimitrova, and H.~Kantz.
\newblock Modified method of simplest equation and its application to nonlinear
  {PDE}s.
\newblock {\em Applied Mathematics and Computation}, 216:2587--2595, 2010.

\bibitem{Vitanov2010c}
N.~K. Vitanov.
\newblock Modified method of simplest equation: powerful tool for obtaining
  exact and approximate traveling-wave solutions of nonlinear {PDE}s.
\newblock {\em Communications in Nonlinear Science and Numerical Simulation},
  2010.

\bibitem{Bruno2004a}
A.D. Bruno.
\newblock Asymptotic behaviour and expansions of solutions of an ordinary
  differential equation.
\newblock {\em Russian Math. Surveys}, 59(3):429--480, 2004.

\bibitem{Kudryashov2010}
N.~A. Kudryashov and D.~I. Sinelshchikov.
\newblock Nonlinear waves in liquids with gas bubbles with account of viscosity
  and heat transfer.
\newblock {\em Fluid Dynamics}, 45(1):96--112, 2010.

\end{thebibliography}

\end{document}